\documentclass[pra,twocolumn,eqsecnum,superscriptaddress,floatfix]{revtex4}

\usepackage{graphicx} 
\usepackage{color} 
\usepackage{amsmath} 
\usepackage[urlcolor=blue, hyperindex, colorlinks, bookmarks=true,linkcolor=black,citecolor=black]{hyperref} 

\newcommand{\dg}{^\dagger}

\newcommand{\bra}[1]{\langle{#1}|}
\newcommand{\ket}[1]{|{#1}\rangle}

\newcommand{\szo}{{\sigma}^z}

\newcommand{\smo}{{\sigma}^-}
\newcommand{\spo}{{\sigma}^+}
\newcommand{\ano}{{a}}
\newcommand{\cro}{{a}\dg}

\begin{document}

\title{Quantum trajectory equation for multiple qubits in circuit QED: Generating entanglement by measurement}
\date{\today}
\author{Chantal L. Hutchison}
\affiliation{Institute for Quantum Computing and Department of Applied Mathematics, University of Waterloo, Waterloo, Ontario, Canada, N2L 3G1}
\author{J. M. Gambetta}
\affiliation{Institute for Quantum Computing and Department of Physics and Astronomy, University of Waterloo, Waterloo, Ontario, Canada, N2L 3G1}
\author{Alexandre Blais}
\affiliation{D\'epartement de Physique et Regroupement Qu\'eb\'ecois sur les Mat\'eriaux de Pointe, Universit\'e
de Sherbrooke, Sherbrooke, Qu\'ebec, Canada, J1K 2R1}
\author{F. K. Wilhelm}
\affiliation{Institute for Quantum Computing and Department of Physics and Astronomy, University of Waterloo, Waterloo, Ontario, Canada, N2L 3G1}
	  
\begin{abstract}
In this paper we derive an effective master equation and quantum trajectory equation for multiple qubits in a single resonator and in the large resonator decay limit. We show that homodyne measurement of the resonator transmission is a weak measurement of the collective qubit inversion. As an example of this result, we focus on the case of two qubits and show how this measurement can be used to generate an entangled state from an initially separable state. This is realized without relying on an entangling Hamiltonian. We show that, for {\em current} experimental values of both the decoherence and measurement rates, this approach can be used to generate highly entangled states. This scheme takes advantage of the fact that one of the Bell states is decoherence-free under Purcell decay.
\end{abstract}
\maketitle

\section{Introduction} 
\label{sec:indroduction}

In the last few years circuit quantum electrodynamics (circuit QED) \cite{Blais:2004a,Wallraff:2004a,Chiorescu:2004a}, a solid state analog of cavity QED \cite{Haroche:1989a,Haroche:2006a,Mabuchi:2002a}, has grown into a mature field.   This system is based on superconducting qubits \cite{QUBITS} acting as artificial atoms and a distributed
\cite{Blais:2004a,Wallraff:2004a,Schuster:2007a,Houck:2007a,Hofheinz:2008a,Wang:2008a,Leek:2007a,Majer:2007a,Sillanpaa:2007a,Astafiev:2007a,Schuster:2005a,Gambetta:2006a} or lumped \cite{Chiorescu:2004a,Johansson:2006a} resonator which acts as a harmonic oscillator.  Examples of its success are the observation of the particle-like nature of microwave photons \cite{Schuster:2007a}, generation of a single photon \cite{Houck:2007a} and Fock states \cite{Hofheinz:2008a,Wang:2008a}, observation of  Berry's phase \cite{Leek:2007a}, use of a quantum bus to couple qubits \cite{Majer:2007a,Sillanpaa:2007a}, single artificial-atom lasing \cite{Astafiev:2007a}, and observation of the fundamental limit that exists between measurement and dephasing \cite{Schuster:2005a,Gambetta:2006a}.

Evolution of this system obeys a master equation (ME) which has an interaction described by the Jaynes-Cummings Hamiltonian \cite{Jaynes:1963a} and decoherence processes which are well described by  Markovian environments \cite{Gardiner:2004b}. Measurement in this system is done by operating in the dispersive limit (where the detuning between the resonator and the qubit is much larger then their coupling strength).  In this limit, the interaction induces a qubit-state dependent frequency shift on the resonator. By measuring the resonator output voltage with a homodyne measurement, information about the qubit state is obtained. This measurement is a weak continuous-in-time measurement of the quadrature of the resonator and the evolution of the conditional state is described by a quantum trajectory equation (QTE) \cite{Gardiner:2004b,Gambetta:2008a}. The presence of the resonator can make the Hilbert space needed for simulation of the ME or QTE quite large and impractical (especially for a many qubit system). 

It has been shown by us in previous work \cite{Gambetta:2008a} that for a single qubit a polaron transformation can be used to eliminate the resonator dynamics from both the ME and QTE. The resultant ME and QTE have an extra decay channel which represents measurement induced dephasing.  Moreover, measurement is found to be a weak measurement of the qubit inversion operator $\sigma_z$. This transformation has the advantage of being exact for the average evolution.  Extending to more than one qubit is non-trivial and will be discussed in a future publication~\cite{YOUMESOMEONE}. 
Here we follow a simpler approach which yields very accurate results in the large resonator damping case. This large damping limit is particularly useful because it corresponds to a good qubit measurement. We use an adiabatic approximation, similar to that presented in Refs.~\cite{Wiseman:1994c,Doherty:1999a,FRANK1,FRANK2}, to obtain an effective ME and QTE.  From these equations, we find that in this limit homodyne measurement of the resonator corresponds to a measurement of the collective qubit inversion ($\sum_i\delta_i\szo_i$, where $\delta_i$ is a dimensionless parameter determined by the system parameters).

As an application of the derived many-qubit QTE, we consider the case of two qubits and tune the system parameters such that $\delta_1=\delta_2$. In this situation, a measurement has three possible outcomes. For one of these outcomes,  measurement cannot distinguish between an excitation being in either qubit.  As a result, starting with a particular separable input state this measurement will generate a maximally entangled state. Similar ideas were used with trapped ions to generate entanglement~\cite{Duan:2001a,Maunz:2007a}. Theoretical work towards realization of these ideas in circuit QED was already done in Ref.~\cite{sarovar:2005a}.  Here however, we consider realistic decoherence rates as measured in recent experiments~\cite{Schreier:2008a}.  Moreover, to obtain large concurrences our scheme takes advantage of the fact that one of the Bell states is decoherence-free under Purcell decay~\cite{Purcell:1946a}.

The paper is organized as follows. In Sec.~ \ref{sec:label} we derive the underlying ME for many qubits coupled to a common resonator and adiabatically eliminate the resonator degrees of freedom. In Sec.~\ref{sec:QT} we derive the corresponding QTE.  In Sec.~\ref{sec:generating_entangled_states_by_measurement}  we investigate how entanglement can be generated by measurement and show that, with realistic parameters, high concurrence can easily be reached. We summarize our findings in Sec.~\ref{sec:conclusion}. 

		    
\section{The Master Equation}\label{sec:label}
   
\begin{figure}[htbp]
	\centering
		\includegraphics[width=0.45\textwidth]{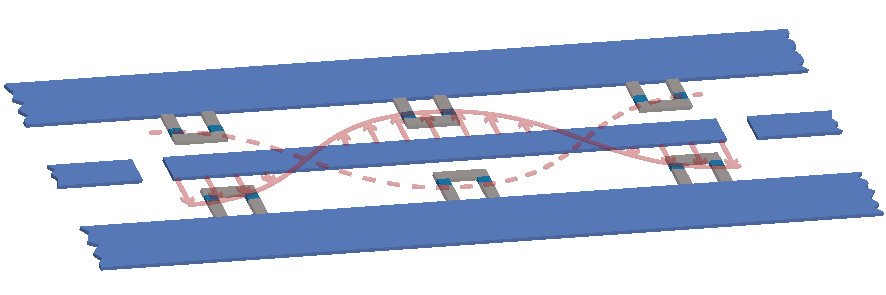}
	\caption{A schematic illustration of multiple-qubit circuit QED. Superconducting qubits (gray; six qubits are illustrated) are fabricated inside a transmission-line resonator (blue).  The full-wavelenght mode of the resonator is illustrated (pink).}
	\label{fig:SlavedQTsFig1}
\end{figure}

We consider multiple superconducting qubits coupled to a
transmission line resonator acting as a simple harmonic
oscillator. This system is illustrated schematically in
Fig.~\ref{fig:SlavedQTsFig1}. In the limit where direct capacitive qubit-qubit coupling can be ignored, the system is described by the multi-qubit Jaynes-Cummings Hamiltonian \cite{Jaynes:1963a,Tavis:1968a} ($\hbar=1$), 
\begin{equation}\label{eq:JaynesCHam}
\begin{split}
		H =& \omega_r\cro\ano+\mathcal{E}(\ano+\cro)+\sum_j\left[\frac{ \omega_{q_j}}{2}\szo_j +  g_j (\smo_j\cro+\spo_j\ano)\right].
\end{split}
\end{equation} 
In this expression, $\omega_r$ is the resonator frequency, $\omega_{q_j}$ the $j^\mathrm{th}$-qubit
transition frequency, and $g_j$ the $j^\mathrm{th}$-qubit-resonator coupling
strength. $\mathcal{E}=\mathcal{E}_m+\mathcal{E}_c$ represents external driving of the resonator, with $\mathcal{E}_m$ referring to the measurement drive (at frequencies close to the resonator) and $\mathcal{E}_c$ referring to the control drives (at frequencies close to the qubits). We take
\begin{equation}
\begin{split}
		\mathcal{E}_m&= \epsilon_m e^{-i\omega_m t}+\epsilon^*_m e^{i\omega_m t}\\
		\mathcal{E}_c&=\sum_{k} \left(\epsilon_k e^{-i\omega_k t}+\epsilon^*_k e^{i\omega_k t}\right),
\end{split}
\end{equation} with $\omega_m$ being close to $\omega_r$ and where, for all $k$, $\omega_k$ is far detuned from the resonator.

In the dispersive regime, where $|\Delta_j|=|\omega_{q_j}-\omega_r|\gg |g_j|$, we can adiabatically eliminate the resonant Jaynes-Cummings interaction using the transformation \footnote{This is a simple extension Eqs. (5.1) and (5.2) in Ref.~\cite{Blais:2007a} to include more then two qubits. To include higher order effects the results of Ref.~\cite{Boissonneault:2008b} could be used.} 
\begin{equation}
	\mathbf{U}= \exp\Big{[}\sum_j\lambda_j(\cro\smo_j-\ano\spo_j)\Big{]},
\end{equation} where $\lambda_j=g_j/\Delta_j$  is a small parameter. The effective Hamiltonian $H_\mathrm{eff} = \mathbf{U}\dg H\mathbf{U}$ is, to second order in $\lambda_j$,
\begin{equation}
\begin{split}\label{eq:dispersiveHam}
	H_\mathrm{eff} = & \omega_r a\dg a +  \mathcal{E}_m(a+a\dg)+\sum_{j>k} J_{jk}(\smo_j\spo_k+\spo_j\smo_k)\\
	&+ \sum_j\left[\chi_j a\dg a \sigma^z_j  + \frac{\omega_{q_j}+\chi_j}{2}\sigma^z_j  + \lambda_j\mathcal{E}_c( \spo_j+ \smo_j)\right],
\end{split}\end{equation} where $\chi_j=g_j^2/\Delta_j$ parametrizes the strength of the ac-Stark shift (fourth term) and Lamb shift (fifth term) on the $j^\mathrm{th}$ qubit transition frequency.  $J_{jk}$ represents the strength of the coupling between the $j^\mathrm{th}$ and $k^\mathrm{th}$ qubit by virtual photons in the resonator and is given by 
\begin{equation}
	J_{jk}=\frac{g_jg_k(\Delta_j+\Delta_k)}{2\Delta_j\Delta_k}.
\end{equation} 

Coupling to additional uncontrolled degrees of freedom
leads to energy relaxation and dephasing in the system.
Integrating out these degrees of freedom leaves the qubit
plus resonator system in a mixed state 
$\rho(t)$ whose evolution can
be described by the ME
\begin{equation}\label{eq:master}
	\begin{split}
		\dot \rho =& -i [H_\mathrm{eff},\rho]+ \kappa {\cal D}[a]\rho + \sum_j \gamma_{1,j} {\cal D}[\sigma^-_j] \rho \\&+ \sum_j\frac{\gamma_{\phi_j}}{2} {\cal D}[\sigma_j^z]\rho+\kappa\mathcal{D}\left[\sum_j\lambda_j\smo_j\right]\rho,
	\end{split}
\end{equation} where 
\begin{equation}
	{\cal D}[A]\rho = A\rho A\dg - \{A\dg A,\rho\}/2.
\end{equation} 
In this expression, $\kappa$ is the resonator decay rate, $ \gamma_{1,j}$, $\gamma_{\phi_j}$ represents relaxation and dephasing on the $j^\mathrm{th}$ qubit and the last term represents correlated relaxation due to the Purcell effect \cite{Purcell:1946a,Boissonneault:2008b,Houck:2008a}. While this terms is of order $\lambda^2$, it is kept as in the following adiabatic approximation we will require $\kappa$ to be large such that the product $\lambda^2\kappa$ is not necessarily small.

To derive an effective ME for the qubits only, we start by removing the fast dynamics of the resonator. This is done by moving to the interaction frame rotating at the measurement drive frequency, $\omega_m$ and by making the standard rotating-wave approximation. This allows us to rewrite Eq.~\eqref{eq:dispersiveHam} as
\begin{equation}
	H_\mathrm{eff} = \Delta_r a\dg a + \sum_j\chi_j a\dg a \sigma^z_j  +  (\epsilon_m^*a+\epsilon_m a\dg) +H_q,
\end{equation}
where $\Delta_r=\omega_r-\omega_m$ and $H_q$ represents the
Hamiltonian of the isolated qubits
\begin{equation}
	\begin{split}
	H_q=& \sum_j\frac{\omega_{q_j}+\chi_j}{2}\sigma^z_j+\sum_{j>k} J_{jk}(\smo_j\spo_k+\spo_j\smo_k)\\&+ \sum_{j}\lambda_j\mathcal{E}_c( \spo_j+ \smo_j).
	\end{split}\end{equation}

Next, we move to the frame defined by 
\begin{equation}
	\rho^\mathbf{D}(t)=\mathbf{D}\dg[\alpha]\rho(t)\mathbf{D}[\alpha],
\end{equation}
where $D[\alpha]=\exp[\alpha a\dg -\alpha^*a]$ is the displacement operator. Applying this to Eq.~\eqref{eq:master} and choosing \begin{equation}
	\alpha=-\frac{i\epsilon_m}{i\Delta_r+\kappa/2}
\end{equation}  
yields
\begin{equation}
	\begin{split}
		\dot \rho^\mathbf{D}=& \mathcal{L} \rho^\mathbf{D}-i\Delta_r[\cro\ano,\rho^\mathbf{D}] +\kappa {\cal D}[a]\rho^\mathbf{D}-i\sum_j\chi_j[a\dg a\sigma_j^z,\rho^\mathbf{D}] \\&-i\sum_j\chi_j[(\alpha^*a+\alpha a\dg)\sigma_j^z,\rho^\mathbf{D}],
	\end{split}
\end{equation} where $\mathcal{L}$ is the Lindblad superoperator representing only qubit dynamics and is given by 
\begin{equation}
\begin{split}
		\mathcal{L}\rho =&-i [H_q,\rho] -i\bar\chi|\alpha|^2\sum_j [\delta_j\sigma_j^z,\rho] + \sum_j \gamma_{1,j} {\cal D}[\sigma^-_j] \rho \\&+ \sum_j\frac{\gamma_{\phi_j}}{2} {\cal D}[\sigma_j^z]\rho+\kappa\mathcal{D}\left[\sum_j\lambda_j\smo_j\right]\rho.
\end{split}
\end{equation} In this equation the second term represents the ac-Stark shift on the qubit transition frequency, with  $\bar\chi=\sum_j\chi_j/N$ and $N$ being the total number of qubits. $\delta_j=\chi_j/\bar\chi$ is the fractional effect that $\chi_i$ gives to the average. 

Following Refs.~\cite{Wiseman:1994c,Doherty:1999a}, we now make an adiabatic approximation. That is, we will assume that quantum fluctuations in the displaced resonator state are small.  In this situation, it is reasonable to assume that matrix elements $\rho_{nm}$, with $n,m$ being the resonator photon number, decay rapidly with increasing $n,m$.  As a result, we will assume that there is a small parameter $\varepsilon$ such that $\rho_{nm} \propto \varepsilon^{n+m}$~\cite{Wiseman:1994c,Doherty:1999a}. Expanding the total state matrix to second order in $\varepsilon$ we find
\begin{equation}
\begin{split}
	\rho^\mathbf{D} =& \rho_{00}\ket{0}\bra{0} + \rho_{10}\ket{1}\bra{0} + \rho_{01}\ket{0}\bra{1} + \rho_{11}\ket{1}\bra{1}\\&
	+\rho_{20}\ket{2}\bra{0}+\rho_{02}\ket{0}\bra{2}+ O(\varepsilon^3)
\end{split}
\end{equation} 
such that the reduced state for the qubits is given by $\varrho=\mathrm{Tr}[\rho^\mathbf{D}]=\rho_{00}+\rho_{11}$. 

Substituting this expansion in the above ME leads to the following coupled differential equations
\begin{equation}\label{eq:slaved}
	\begin{split}
		\dot\rho_{00} =& \mathcal{L} \rho_{00} + \kappa\rho_{11} + i\sum_j\chi_j(\alpha\rho_{01}\sigma_j^z -\alpha^*\sigma_j^z\rho_{10}), \\
		\dot\rho_{10} =& \mathcal{L} \rho_{10} - \kappa\rho_{10}/2 + i\sum_j\alpha\chi_j(\rho_{11}\sigma_j^z -\sigma_j^z\rho_{00})\\& -i\sum_j\chi_j(\sigma^z_j\rho_{10}+\alpha^*\sqrt{2}\sigma^z_j\rho_{20})-i\Delta_r\rho_{10} , \\
		\dot\rho_{11} =& \mathcal{L} \rho_{11} - \kappa\rho_{11} + i\sum_j\chi_j(\alpha^*\rho_{10}\sigma_j^z-\alpha\sigma_j^z\rho_{01}) \\&-i\sum_j\chi_j[\sigma^z_j,\rho_{11}], \\
		\dot\rho_{20} =& \mathcal{L} \rho_{20} - \kappa\rho_{20} -i2\Delta_r\rho_{20}\\& - i\sum_j \chi_j(\alpha\sqrt{2}\sigma_j^z\rho_{10}+ 2\sigma^z_j\rho_{20}).	
	\end{split}
\end{equation}
By looking closely at these expressions, we find that the higher order terms are only populated at rate $|\alpha|\sum_j\chi_j/\kappa$ and decay at rate $\kappa$. Thus, taking
\begin{equation}
	\varepsilon = \sum_j\chi_j|\alpha|/\kappa\ll1
\end{equation} 
as our small parameter, we can assume that the off-diagonal terms $\rho_{10}$ and $\rho_{20}$ decay much faster than the diagonal terms and as such can be approximated by their steady-state value
\begin{equation}\begin{split}\label{eq:rho01}
	\rho_{10} &= \frac{i\alpha\sum_j\chi_j(\rho_{11}\sigma_j^z -\sigma_j^z\rho_{00})}{i\Delta_r+\kappa/2},\\
	\rho_{20} &= \sum_j\frac{-i\alpha\sqrt{2}\chi_j\sigma_j^z \rho_{10}}{\kappa+i2\Delta_r}.
\end{split}
\end{equation}
Substituting these expressions into the diagonal components leads to the effective ME \footnote{Note that, as in Ref.~\cite{Wiseman:1994c,Doherty:1999a}, we have kept $\rho_{11}$ even though to the required order it can be left out.}
\begin{equation}\label{eq:SlavedMasterEq}
	\dot\varrho = {\cal L}\varrho  + \frac{\Gamma_\mathrm{d}}{2}{\cal D}\left[\sum_j\delta_j\sigma_j^z\right]\varrho -iK[(\sum_j \delta_j \szo_j)^2,\varrho].
\end{equation} 
In this expression, $\Gamma_\mathrm{d}$ is the measurement-induced dephasing rate and $K$ is a resonator-induced Ising-like coupling. These are 
\begin{equation}
	\Gamma_\mathrm{d} = \frac{2\kappa |\alpha|^2\bar\chi^2}{\Delta_r^2+\kappa^2/4}~\mathrm{and}~ K=\frac{-2\Delta_r|\alpha|^2\bar\chi^2}{\Delta_r^2+\kappa^2/4}.
\end{equation} 

In the limit of a single qubit, the measurement
induced-dephasing rate obtained here correctly agrees with the large $\kappa$ limit of
Ref.~\cite{Gambetta:2006a} (In this paper, this expression is labeled
$\Gamma_\mathrm{m}$) and with the large $\kappa$ and steady-state
limit of the rate presented in Ref.~\cite{Gambetta:2008a}. We note that one could use a
multi-qubit polaron transformation to get results valid
in the small $\kappa$ limit and which would take into account initial transients
in the resonator~\cite{YOUMESOMEONE}.

\section{Quantum trajectory equation}\label{sec:QT}

Although direct detection of the transmitted
microwave photons is possible \cite{Houck:2007a}, here we will
consider homodyne processing \cite{Gardiner:2004b}.  That is, we will assume that the signal
coming from the output port of the resonator is mixed with a strong
local oscillator tuned to the signal frequency and of phase $\phi$. Given the homodyne measurement result $J(t)$, we can assign to the qubit and the resonator the conditional state $\rho_J(t)$ whose evolution is governed by the QTE \cite{Wiseman:1993a}
\begin{equation}
\label{eq:HomodyneQT}
		\dot\rho_J
		={\cal L}\rho_J
		+i\sqrt{\kappa\eta}[Q_\phi,\rho_J]\xi(t)+\sqrt{\kappa\eta}{\cal M}[2 I_\phi]\rho_J\xi(t),
\end{equation}
with ${\cal L}$ given above. Here, ${\cal M}[{ c}]$ is the measurement superoperator defined as
 \begin{equation}
	{\cal M}[{c}]\rho= ({c} -\langle{c}\rangle_t) \rho/2+\rho({c}-\langle{c}\rangle_t)/2,
\end{equation}
where $\langle{c}\rangle_t = \mathrm{Tr}[ c \rho_J(t)]$ and
the $\phi$-dependent field components are $2I_\phi = a e^{-i\phi} + a \dg e^{i\phi}$ and $2 Q_\phi = -i a e^{-i\phi} + i a \dg e^{i\phi}$.  $\eta$ is the  efficiency at which the photons coming out of the resonator are detected.  For the current circuit QED experiments, this can be written as $\eta_\mathrm{det} = 1/(N_\mathrm{th}+1)$ with $N_\mathrm{th}$ being the number of noise photons added in the amplifier stage. The measurement record is
\begin{equation}
	J(t) =\sqrt{\kappa\eta} \langle2 I_\phi \rangle_t +\xi(t),
\end{equation}
where $\xi(t)$ is Gaussian white noise and represents the photon shot noise. It is formally defined by ${\rm E}[\xi(t)]=0$ and ${\rm E}[\xi(t)\xi(t')]=\delta(t-t')$,
with ${\rm E}$ denoting an ensemble average over realizations of the noise $\xi(t)$.

To obtain an effective QTE for the qubits only, we apply the small $\varepsilon$ expansion to the above QTE.  For the stochastic part only, we find
\begin{equation}
	\begin{split}
		\dot\rho_{00} =&\sqrt{\kappa\eta}\left(\rho_{10}e^{-i\phi}+\rho_{01}e^{i\phi}-\langle2I_\phi\rangle\rho_{00} \right)\xi(t),\\
			\dot\rho_{11} =&\sqrt{\kappa\eta}\left(-\langle2I_\phi\rangle\rho_{11} \right)\xi(t),
	\end{split}
\end{equation} which gives the following stochastic term to the qubit equation
\begin{equation}
	\dot\varrho = \sqrt{\kappa\eta}\left(\rho_{10}e^{-i\phi}+\rho_{01}e^{i\phi}-\langle2I_\phi\rangle\varrho \right)\xi(t).
\end{equation}
Using the steady state value for $\rho_{10}$ given in Eq. \eqref{eq:rho01} this yields 
\begin{equation}
	\dot\varrho \approx \sqrt{\frac{4 \kappa\eta}{\Delta_r^2+\kappa^2/4}}\bar\chi\left\{X\mathcal{ M}[\sum_i\delta_i\sigma_i^z] -i\frac{Y}{2}\left [\sum_i\delta_i\sigma_i^z,\varrho\right]\right\}\xi(t),
\end{equation} where $X = \Re[\alpha^* e^{i(\phi-\theta)}]$ and $Y=\Im[\alpha^* e^{i(\phi-\theta)}]$ with $\tan(\theta)=\kappa/2\Delta_r$. Defining the measurement rate $\Gamma_\mathrm{ci}$, the extra non-Heisenberg backaction $\Gamma_\mathrm{ba}$ and the maximum measurement rate $\Gamma_\mathrm{m}$ as
\begin{equation}
	\Gamma_\mathrm{ci}= \frac{4 \kappa\eta\bar\chi^2X^2}{\Delta_r^2+\kappa^2/4},
	~\Gamma_\mathrm{ba}=\frac{4 \kappa\eta\bar\chi^2Y^2}{\Delta_r^2+\kappa^2/4},~\mathrm{and}~
	\Gamma_\mathrm{m}= \frac{4 \kappa\bar\chi^2|\alpha|^2}{\Delta_r^2+\kappa^2/4},
\end{equation} allows us to write the effective QTE in the form
\begin{equation}\label{eq:SlavedQT}
\begin{split}
		\dot\varrho_J =& {\cal L}\varrho_J  + \frac{\Gamma_\mathrm{d}}{2}{\cal D}\left[\sum_j\delta_j\sigma_j^z\right]\varrho_J - iK[(\sum_j \delta_j \szo_j)^2,\varrho_J]\\
		&+\sqrt{\Gamma_\mathrm{ci}}{\cal M}[\sum_i\delta_i\sigma_i^z]\varrho_J\xi(t) -i\frac{\sqrt{\Gamma_\mathrm{ba}}}{2} \left [\sum_i\delta_i\sigma_i^z,\varrho\right]\xi(t)
\end{split}
\end{equation}
with
\begin{equation}
	J(t) =\sqrt{\Gamma_\mathrm{ci}} \sum_i\langle \delta_i\sigma_i^z \rangle_t +\xi(t).
\end{equation} and 
$\Gamma_\mathrm{ci}+\Gamma_\mathrm{ba} = \eta\Gamma_\mathrm{m}$. This last equality is also found in Ref.~\cite{Gambetta:2008a}  and reflects the fact that maximum information about the qubit is obtained by setting the phase of the local oscillator such that $\Gamma_\mathrm{ba}$ is zero ($Y=0$). At this point, the rate of information gain is $\Gamma_\mathrm{ci}=\eta\Gamma_\mathrm{m}$ and for $\eta=1$ this system reaches the quantum limit ($\Gamma_\mathrm{d}/\Gamma_\mathrm{ci}=1/2$).

\section{Entanglement by measurement} 
\label{sec:generating_entangled_states_by_measurement}

\begin{figure}[htbp]
	\centering
		\includegraphics[width=0.45\textwidth]{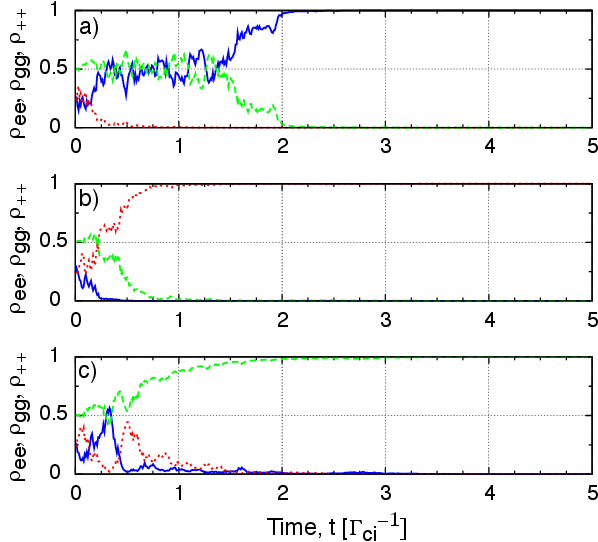}
		\caption{Three typical trajectories of the conditional state elements $\rho_{ee}$ (blue solid line) $\rho_{gg}$ (red dotted line) and $\rho_{++}$ (green dashed line) when the initial condition is the separable state given by Eq.~\eqref{eq:sepinitialstate}. The system parameters are  $\Gamma_\mathrm{d}=\Gamma_\mathrm{ci}/2$, and $\gamma_{1,1}=\gamma_{1,2}=\gamma_p=0$. In panel a), the collapse of the wavefunction is to the pure state $\ket{ee}$, in panel b) to $\ket{gg}$ and in panel c) to the entangled state $\ket{+}$.}
	\label{fig:SlavedQTsFig2}
\end{figure}

As an illustration of application of the above QTE, we consider the probabilistic generation of entanglement from a separable state without using a two-qubit unitary. This approach can be particularly useful when such a two-qubit unitary is not present or hard to implement. We consider the case where only two qubits are present and take $\Delta_r=0$, $\delta_1=\delta_2=1$. The last equality is such that both states $|eg\rangle$ and $|ge\rangle$ pull the resonator to the same frequency and are thus indistinguishable.  Moreover, the local oscillator phase is assume to be chosen such that $\Gamma_\mathrm{ba}=0$ and we take $\lambda_1=-\lambda_2$.  The sign change in $\lambda$ is easily realized by putting the two qubits at both ends of the resonator and working with an odd mode.   Finally, to rule out any possible entanglement due to unitary qubit-qubit coupling the indirect interaction $J_{12}$ between the qubits is ignored in the numerics~\footnote{Keeping this coupling does not change the results since the states generated by measurement are eigenstates of the $J_{12}$ interaction.}.

With a transmon-type qubit~\cite{Koch:2007a}, we can safely take the qubit dephasing rate $\gamma_\phi$ to be small \cite{Schreier:2008a}. In this situation, Eq.~\eqref{eq:SlavedQT} in the rotating frame becomes
\begin{equation}
		\begin{split}
			\dot\varrho_J =&\gamma_{1,1}\mathcal{D}[\smo_1]\varrho_J+\gamma_{1,2}\mathcal{D}[\smo_2]\varrho_J +\gamma_p\mathcal{D}[\smo_1 -\smo_2]\varrho_J \\&+ \frac{\Gamma_\mathrm{d}}{2}{\cal D}\left[\sigma_1^z+\sigma_2^z\right]\varrho_J +\sqrt{\Gamma_\mathrm{ci}}{\cal M}[\sigma_1^z+\sigma_2^z]\varrho_J\xi(t),
		\end{split}
\end{equation}
where $J=\sqrt{\Gamma_\mathrm{ci}}\langle \szo_1+\szo_2\rangle + \xi(t)$ and $\gamma_\mathrm{p}=\kappa \lambda^2$ is the Purcell decay rate.

For the system in the initial separable state 
\begin{equation}\label{eq:sepinitialstate}
	\ket{\psi}=\frac{1}{2}(\ket{e}+\ket{g})(\ket{e}+\ket{g})=\frac{1}{2}(\ket{ee}+\ket{gg}+\sqrt{2}\ket{+}),
\end{equation}
where \begin{equation}
	\ket{\pm}= \frac{1}{\sqrt{2}}\left(\ket{eg}\pm\ket{ge}\right),
\end{equation} 
the QTE (for $\gamma_{1,1}=\gamma_{1,2}=\gamma_p=0$) leads to  a collapse to the state $\ket{ee}$ with probability $1/4$, $\ket{gg}$ with probability $1/4$ and to the entangled state $\ket{+}$ with probability $1/2$. This is illustrated in Fig.~\ref{fig:SlavedQTsFig2} where typical quantum trajectories for the elements $\rho_{ee}$ (blue solid) $\rho_{gg}$ (red dotted) and $\rho_{++}$ (green dashed) are plotted as a function of time for these three possible outcomes. 

\begin{figure}[htbp]
	\centering
		\includegraphics[width=0.45\textwidth]{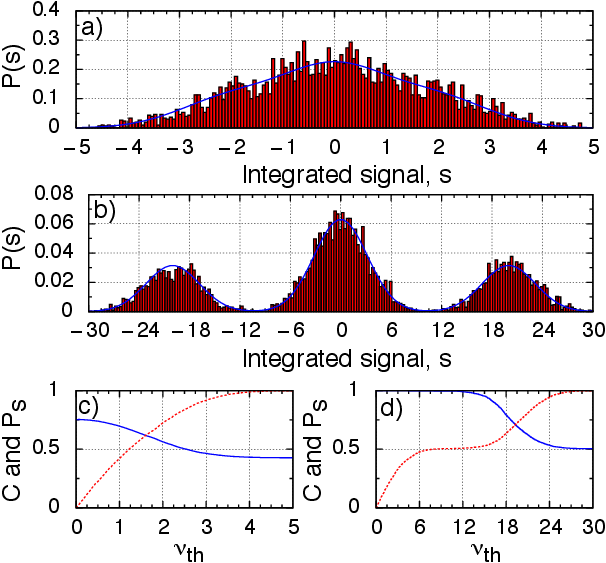}
\caption{Histogram of 10000 trajectories at time $t=1\Gamma_\mathrm{ci}^{-1}$ a) and time $t=10\Gamma_\mathrm{ci}^{-1}$ b). In both panels the full blue line shows the expected distribution.  The center peak in panel b) corresponds to the entangled state $\ket{+}$ while the side peaks to the separable states $\ket{gg}$ and $\ket{ee}$. Average concurrence (blue solid line) and probability of success (red dotted line) are plotted as a function of $\nu_\mathrm{th}$ for times  $t=1\Gamma_\mathrm{ci}^{-1}$ c) and  $t=10\Gamma_\mathrm{ci}^{-1}$ d). Other parameters are the same as in Fig.~\ref{fig:SlavedQTsFig2}.}
	\label{fig:SlavedQTsFig3}
\end{figure}

To connect this to the language of POVM's \cite{Nielsen:2000b}, or to a quantity measured in an experiment, we defined the integrated current \cite{Gambetta:2007a,Gambetta:2008a}
\begin{equation}
	s(t) = \int_0^t J(t') dt'.
\end{equation} If we break $s$ into three regimes defined by $s<-\nu_\mathrm{th}$, $-\nu_\mathrm{th}<s<\nu_\mathrm{th}$, and $s>\nu_\mathrm{th}$, where we refer to $\nu_\mathrm{th}$ as the threshold then the POVM elements $E_g$, $E_0$, and $E_e$ are measured, respectively. These are
\begin{equation}
\begin{split}
	E_{g}&=a_{g} \Pi_{g} +b_{g}\Pi_{0}+c_{g}\Pi_{e}, \\
	E_{0}&=a_{0} \Pi_{g} + b_{0}\Pi_{0}+c_{0}\Pi_{e},\\
	E_{e}&=a_{e} \Pi_{g} + b_{e}\Pi_{0}+c_{e}\Pi_{e}.
\end{split}
\end{equation} where $\Pi_{g}=\ket{gg}\bra{gg}$, $\Pi_{e}=\ket{ee}\bra{ee}$ and $\Pi_{0}= \ket{eg}\bra{eg}+\ket{ge}\bra{ge}$ are projectors and the rest of the parameters are simply positive real numbers (probabilities) which satisfy  $a_{g}+a_{0}+a_{e}=1$, $b_{g}+b_{0}+b_{e}=1$, and $c_{g}+c_{0}+c_{e}=1$. For example in the POVM element $E_g$, $a_{g}$ is the probability that the measurement was of the projector we desired and $b_g$ and $c_g$ are the probability of the false positive events $\Pi_0$ and $\Pi_e$ respectively.  

A histogram of $s$ for 10000 trajectories is plotted in Fig. \ref{fig:SlavedQTsFig3} for $\gamma_{1,1}=\gamma_{1,2}=\gamma_p=0$. Panel a) is taken at the integration time $t=1\Gamma_\mathrm{ci}^{-1}$ and panel b) at $t=10\Gamma_\mathrm{ci}^{-1}$. Clearly, the measurement at the earlier time is not projective (the false positive rates are high), whereas at the later time the measurement is projective since the distributions are well separated. At that time, it is possible to create and distinguish the entangled state $\ket{+}$ from $\ket{gg}$ and $\ket{ee}$. This is shown more explicitly in panels c) and d) where the average concurrence~\cite{Wootters:1998a} and the probability of success $P_s$ (defined as the number of detections in the range $[-\nu_\mathrm{th},\nu_\mathrm{th}]$ divided by total number of measurements)  as a function of the threshold $\nu_\mathrm{th}$ are plotted. Panel c) corresponds to $t=1\Gamma_\mathrm{ci}^{-1}$ and panel d) to $t=10\Gamma_\mathrm{ci}^{-1}$.  We see that, at the shorter time, it is impossible to generate the entangled state while at the later time a highly entangled state can be created by setting a low threshold. If this threshold is too small the probability of generating the entangled state goes to zero and if it is too large then the unentangled results are included in the post selection states. However, if the measurement time is long enough there can be a large range of values for $\nu_\mathrm{th}$ [8--12 in Fig.~\ref{fig:SlavedQTsFig3} d)] where it is possible to create the desired entangled state with probability $1/2$. 

\begin{figure}[htbp]
	\centering
		\includegraphics[width=0.45\textwidth]{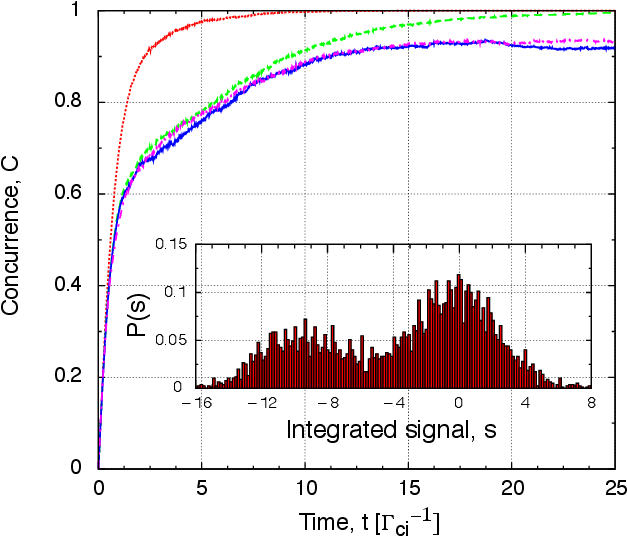}
	\caption{The average concurrence given that $s$ is between $\pm\sqrt{\Gamma_\mathrm{ci}}t$  as a function of time for 5000 trajectories and  $(\gamma_\mathrm{p},\gamma_{1,1},\gamma_{1,2},\Gamma_\mathrm{d}) = (0,0,0,1/2)\Gamma_\mathrm{ci}$ (red dotted line), (1/2,0,0,1/2)$\Gamma_\mathrm{ci}$ (green dashed line), (1/2,1/200,1/200,1/2)$\Gamma_\mathrm{ci}$ (blue solid line),  and (1/2,1/200,1/200,10)$\Gamma_\mathrm{ci}$ (pink dash-dotted line). The inset is a histogram of 5000 trajectories at time $t=5\Gamma_\mathrm{ci}^{-1}$ and $(\gamma_\mathrm{p},\gamma_{1,1},\gamma_{1,2},\Gamma_\mathrm{d}) =(1/2,0,0,1/2)\Gamma_\mathrm{ci}$.}
	\label{fig:SlavedQTsFig4} 
\end{figure}

We now take into account realistic values for the qubits relaxation rate. We choose $\gamma_\mathrm{p}=\Gamma_\mathrm{ci}/2$ and $\gamma_{1,1}=\gamma_{1,2}=\Gamma_\mathrm{ci}/200$ as measured in Ref.~\cite{Houck:2008a}. These rates correspond  to $T_p = 1/\gamma_p= 40$ ns and $T_1 =  4$ $\mu$s, and to a signal-to-noise ratio $\Gamma_\mathrm{ci}/\gamma_p$ of 2 consistent with experimental observations. Numerical results using these values presented in Fig.~\ref{fig:SlavedQTsFig4} show that states that are close to maximally entangled can be obtained even in the presence of qubit decay. This is because the Purcell effect, which is the dominating source of decay for the transmon~\cite{Houck:2008a}, does not act on the entangled state $\ket{+}$, since $(\smo_1 -\smo_2)\ket{+}=0$.  This state is thus a decoherence-free subspace with respect to this decay channel.  It is also worth pointing that it is an eigenstate of the Hamiltonian, such that the prepared $\ket{+}$ states are robust against further evolution of the system.

To maximize the amount of entanglement generated, we find that it is best to use an integration time which is larger then $1/\gamma_\mathrm{p}$ (so that errors due to to $\ket{ee}$ decaying into $\ket{-}$ have subsided) but shorter than the single qubit relaxation time $1/\gamma_\mathrm{1,1(2)}$ (causing errors with $\ket{+}$ decaying into $\ket{gg}$).  Doing this, we find that the maximum attainable concurrence can be as large as $0.94$.  Interestingly, these results depend on the signal-to-noise ratio $\Gamma_\mathrm{ci}/\gamma_p$, but not on the efficiency of the detector $\eta$ which was so far taken to be unity.  To show this, we have simulated the average concurrence for $s$ between $\pm \sqrt{\Gamma_\mathrm{ci}}t$ as a function of time and for an efficiency $\eta=1/20$ ($\Gamma_\mathrm{d}=10\Gamma_\mathrm{ci}$). This is shown in Fig.~\ref{fig:SlavedQTsFig4}c) as the pink dash-dotted line.  These results are equal, within statistical error, to the situation where $\eta=1$ (blue solid line).  This is simply because the extra dephasing caused by the inefficiency only results in the system losing coherences between the measurement eigenstates faster than the limit imposed by measurement (the quantum limit). This does not change the measurement statistics as, with dephasing, measurement selects the final state out of a classical mixture rather than from a quantum superposition.

\section{Conclusion} 
\label{sec:conclusion}

In this paper we have derived a master and quantum trajectory equation for multiple qubits in a resonator. These equations are valid in the limit where $\varepsilon = \sum_j\chi_j|\alpha|^2/\kappa \ll 1$. In this limit, we find that measurement of the resonator transmission leads to a weak measurement of the qubit observable $\sum_i\delta_i\szo_i$.  As an example of this result, we have discussed how entanglement can probabilistically be generated by measurement only.  We have shown that with current experimental decoherence rates, this process can yield a highly entangled state.  This is mostly due to the fact that, in circuit QED, a major source of relaxation (the Purcell effect) does not affect one of the Bell states which is therefore protected from decay.

\begin{acknowledgments}
We thank Jens Koch for discussions.
CLH and FKW acknowledge support by NSERC through their discovery grants
program, Quantumworks, and the USRA program.
JMG was supported by CIFAR, MITACS and ORDCF.
AB was supported by NSERC, FQRNT and CIFAR.
\end{acknowledgments}

\end{document}